\documentclass[]{aa}
\usepackage[dvipdfmx]{graphicx}
\usepackage{txfonts}
\usepackage{natbib}

\begin{document}

\title{On the nature of transverse coronal waves revealed by wavefront dislocations}
\author{{\sc A.~L{\'o}pez Ariste}\inst{1,2}, M. Luna\inst{3,4}, I. Arregui \inst{3,4}, E. Khomenko\inst{3,4}, M. Collados\inst{3,4}}
\institute{THEMIS - CNRS UPS 853. C/ V\'{\i}a L\'actea s/n. 38200. La Laguna. Spain.\and IRAP - CNRS UMR 5277. 14, Av. E. Belin. 31400 Toulouse. France \and Instituto de Astrof\'{\i}sica de Canarias. C/ V\'{\i}a L\'{a}ctea s/n. 
38200 La Laguna. Spain. \and
Departamento de Astrof\'{\i}sica, Universidad de La Laguna, 38205 La Laguna, Tenerife, Spain.}
\offprints{Arturo.LopezAriste@irap.omp.eu}
\date{Received ; accepted}
\begin{abstract} 
{Coronal waves are an important aspect of the dynamics of the plasma in the corona. Wavefront dislocations are topological features 
of most waves in nature and \textbf{also} of magnetohydrodynamic waves. Are there dislocations in coronal waves?}
{The finding and explanation of dislocations may shed light on the nature
and characteristics of the propagating waves, their interaction in the corona and in general on the plasma dynamics.}
{We positively identify dislocations in coronal waves observed by the Coronal Multi-channel Polarimeter (CoMP) as singularities in the Doppler shifts of 
emission coronal lines. We study the possible singularities that 
can be expected in coronal waves and try to reproduce the observed dislocations in terms of localization and frequency of appearance.}
{The observed dislocations can only be explained by the interference of a kink and a sausage wave modes propagating with 
different frequencies along the
coronal magnetic field. \textbf{In the plane transverse to the propagation, the cross-section of the oscillating plasma} must be smaller than the spatial resolution, and the two waves result in net longitudinal and transverse 
velocity components that are mixed through projection onto the line of sight. Alfv\'en waves can be responsible of the kink mode, but a magnetoacoustic
sausage mode is necessary in all cases. Higher (flute) modes are excluded. \textbf{The kink mode has a pressure amplitude 
that is smaller than the pressure amplitude of the sausage mode, though its observed velocity is larger.} This concentrates dislocations on the top of 
the loop. }
{\textbf{To explain dislocations, any model of coronal waves must include the simultaneous propagation and interference of kink and sausage wave modes of 
comparable but different  frequencies, with a sausage wave amplitude much smaller than the kink one.} }
\end{abstract}

\authorrunning{L\'opez Ariste et al.}
\titlerunning{Dislocations in coronal waves}
\maketitle

\section{Introduction}
Magnetohydrodynamic transverse waves seem to be a relevant constituent in 
the dynamics of magnetic and plasma structures in the solar atmosphere. 
Their presence has been invoked to explain imaging and spectroscopic signatures 
of periodic plasma motions detected in different types of structures, 
with different physical conditions, such as coronal loops \citep{aschwanden_coronal_1999,nakariakov_trace_1999}; chromospheric spicules and mottles 
\citep{de_pontieu_chromospheric_2007}; soft X-ray coronal jets \citep{cirtain_evidence_2007}; prominence 
fine structures \cite{okamoto_coronal_2007,lin_swaying_2009}; or extended regions of the 
solar corona \citep{tomczyk_alfven_2007}. 
In recent years, their relevance has increased because of their potential as a seismology 
diagnostic tool \citep{arregui_mhd_2007,goossens_analytic_2008,arregui_bayesian_2011,nakariakov_determination_2001,de_moortel_magnetohydrodynamic_2012}  and their 
possible role in wave heating processes \citep{parnell_contemporary_2012,arregui_wave_2015}.

\begin{figure}[htbp]
\resizebox{9cm}{!}{\includegraphics{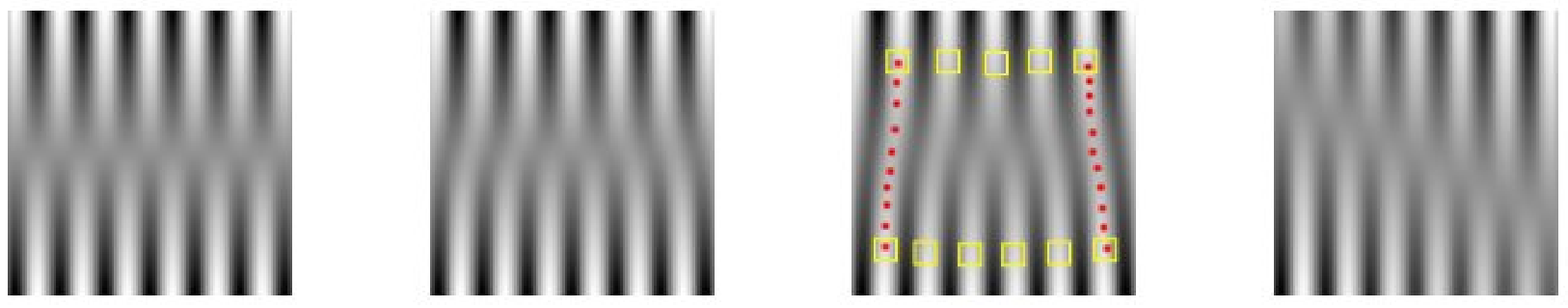}} 
\caption{Four examples of dislocations in the propagation of a wave with the time in abscissas and a distance in ordinates. From left to right, a vortex, two edges and a gliding edge, all with
charge 1. Around the second edge dislocation a closed curve has been drawn with dots and squares. The integral of the phase along this curve, the monodromy, is non-zero.}
\label{ejemplo}
\end{figure}

\cite{lopez_ariste_dislocations_2013} demonstrated the existence of solutions  to the equation of magneto-hydrodynamic waves carrying wavefront dislocations. In 
that work, the equation studied was Eq. 4.14 from \cite{priest_solar_1982} which reads:
\begin{equation}
 \frac{\partial^2 \vec{v}_1}{\partial t^2} = c_S^2\vec{\nabla}(\vec{\nabla}\vec{v}_1)+
\left[\vec{\nabla}\times(\vec{\nabla}\times(\vec{v}_1\times\vec{B}_0))\right]\times\frac{\vec{B}_0}{\mu \rho_0},
\label{eq}
\end{equation}
where $\vec{v}_1$ is the velocity of the plasma, $\vec{B}_0$ the magnetic field, assumed constant, $c_S$ the speed of sound in the medium,
and $\rho_0$ the density. This equation describes linear magnetohydrodynamical waves in a homogeneous, isothermal medium. This equation has as solutions waves 
propagating along the magnetic field that carry
wavefront dislocations \citep{nye_dislocations_1974}, that is, singularities in the phase of the wave. Examples of such waves can be seen in Fig. \ref{ejemplo}. The 
four images show different kinds of dislocations made by varying the parameters $\beta$ and $\delta$ of a generic solution to the longitudinal (that is, along the magnetic field) 
component of the velocity perturbation $v_z$ in Eq. (\ref{eq}) given by \cite{lopez_ariste_dislocations_2013}:
\begin{equation}
 v_z=A\left(kr^me^{im\theta}+\beta e^{i\delta} k (z-ct)\right)e^{ik(z-ct)}.
 \label{firstsol}
\end{equation}
Starting from the left, the illustrated dislocations are a pure vortex ($\beta=0$), two pure edge dislocations ($\beta\neq 0, \delta=\frac{\pi}{2}$)
and a mixture of vortex and edge that can be described as 
a \textit{gliding} dislocation ($\beta\neq 0, \delta=0$). 
The axes of the plot are time and distance, though any other sensitive choice can be made that captures the topology of the wavefront under scrutiny. When looking
at this kind of pictures it is important to agree in that the quasi-periodic variation seen in those time-distance plots is a
sound representation of
the phase of the wave times an amplitude that does not have to be constant over the plot (actually it will be zero at the singularity and only at that point, but 
can take any other value elsewhere). A wave period can be measured with more or less precision and significance by measuring the distance between two crests 
(white color)
or two valleys (black color) in such a plot. If we were staring at a plane wave, the pattern of black and white crests and valleys of the wave would be roughly 
parallel over the plot. We see that this is not the case and we have encircled a particular region in one of the figures in which one crest ends suddenly where 
two valleys merge into
just one. It is evident that the ending point of that crest cannot have a well-defined phase, it is singular. The amplitude of the wave at this point 
should be zero. This is the dislocation. Sometimes it is just a point (pure edge dislocation) but often it is a line of singularities (as in the vortex case on the 
leftmost figure or the gliding dislocation of the rightmost one).
At this point it is important to distinguish a dislocation from mere wave nodes. A node is just a zero of a wave with real amplitude. It is often 
illustrated in one-dimensional plots of waves, often two interfering waves as in standing waves, where we see the amplitude go to zero. But this definition makes 
no mention of the phase of the wave which, despite the zero in amplitude, may still be perfectly well defined. In a dislocation the phase of the wave is singular. 
Thus, dislocations can be seen as nodes, but not all nodes are dislocations. An illustration is offered in our context of MHD waves by a solution of a wave propagating 
along the $z$ direction and which in the transverse plane $(r,\theta)$ has the form $J_1(r)e^{i\theta}$. Each zero of the Bessel function $J_1$ is a node of the 
wave: the amplitude will be zero at those places at all times. But the node at $r=0$ is a dislocation since the phase $\theta$ is undefined or singular at that
point. On the contrary the first zero at $r=3.83$ is a node for which the phase is perfectly well defined and hence it is not a dislocation. Telling apart a 
dislocation from a node with a wave given as a real function of just one variable is not possible. If the wave is written as a complex function
$\rho e^{i\chi}=a+ib$ the dislocation
is found as that place where $a=b=0$ what immediately leaves the phase $\chi=\arctan \frac{b}{a}$ singular and the 
amplitude $\rho=\sqrt{a^2+b^2}$ zero. If the  wave
is seen \textbf{as a function} of two variables (time and one spatial dimension) as in our examples of Fig. \ref{ejemplo} or Fig. \ref{obs}, the qualitative description made above also 
successfully identifies a dislocation and separates it from other nodes.
But this qualitative picture of a dislocation in a time-distance plot of the wavefront or the simple rule of finding the places where both the
imaginary and real parts of the waves are simultaneously zero must be complemented with a more quantitative and \textbf{mathematically rigorous}
definition of the singularity. This is done by drawing a closed curve, called a \textit{monodromy}, along which we integrate an appropriate parameter. In our case
the monodromy is computed over the  phase: if we describe the wave as a map of $\rho e^{i\chi}$, with $\rho$ and $\chi$ real numbers that define
the amplitude and phase respectively at each and every point, the monodromy of interest is
$$\oint _C  d\chi $$
where $C$ represents that closed curve that can be in our example the circle entouring the singularity. The monodromy is strictly zero when there is no singularity
in the area enclosed by $C$, and it is $2\pi m$ if there is a dislocation, with $m$ the topological charge of the singularity. This charge $m$ is 
named to coincide
with the $m$ in Eq. (\ref{firstsol}). And this is in purpose, since we shall verify that the $m$ in that solution defines the charge of the 
generated dislocation.

\cite{lopez_ariste_dislocations_2013} pointed to observations of magnetoacoustic waves in the sunspot umbra by \cite{centeno_spectropolarimetric_2006} where dislocations as the ones
illustrated could be easily identified visually but also by computation of the monodromy. Dislocations are not extraordinary solutions of the wave equation but
actually quite common occurrences, as those observations demonstrated. In the scenario described by Eq. (\ref{eq}) there is a clear axial symmetry given by the 
constant magnetic field. If, for the time being, we restrict ourselves to waves propagating along this $z$ direction it is well-known
\citep{wentzel_hydromagnetic_1979,spruit_propagation_1982,edwin_wave_1983,roberts_wave_1981} that solutions to this
equation are given in terms of families of Bessel functions times an azimuthal dependence $e^{im\theta}$. This angle $\theta$ corresponding to 
the cylindrical
azimuth coordinate is obviously singular at the origin of the coordinate system $r=0$. Those classic solutions to the Eq. (\ref{eq}) carry
therefore a dislocation
at $r=0$ for all the cases with $m\ne 0$. \textbf{We find here a third occurrence of $m$, this time to refer to the azimuthal wave number of the
solutions to the wave equation. It is sufficient to try the monodromy integral over the gradient of the phase to realize that $m$ is exactly the charge
of the dislocation introduced above. And thus, following the usual naming convention, a wave with charge $m=0$ will be called a \textit{sausage} mode, 
while a wave with charge $m=1$ will be called a \textit{kink} mode. Higher values of the charge $m$ are referred to as \textit{flute} modes.}
Finally, it is interesting to notice and stress at this point that any attempt to describe such singularities with a finite 
combination of colinear plane waves or Fourier components will fail.  Dislocations require either the full infinite series of the Fourier 
decomposition or choosing a family of solutions that 
already carries a dislocation in each or most of its components. The Bessel functions times the azimuthal 
dependence $e^{im\theta}$ are one of such families and
an example that we should expect to find in the description of magnetohydrodynamic waves in the solar atmosphere. It is therefore expected 
that dislocations
are observed in waves in the solar atmosphere, though the common description of waves in terms of plane waves have resulted in overlooking them 
because plane waves cannot describe dislocations.

In the present paper we should turn into another observation of magnetohydrodynamic waves, this time in the solar corona. The observations can be seen in
Fig. \ref{obs} and they will be described in terms of dislocations in the next section. Section 3 is a first attempt to describe the observed dislocations in terms of 
generic wave solutions that require the presence of several interferring waves with different frequencies and propagation velocities. The detailed inspection of 
mathematical solutions in Section 4 will show that although most of the waves propagating in coronal tubes carry dislocations, they are not visible in observations 
like the one in Fig. \ref{obs} unless the observed Doppler velocity is projection of both the transverse and the longitudinal velocity. 
This last one has therefore
to be non-zero, what implies the necessary presence of at least one magnetoacoustic mode in the observed waves.

\section{Dislocations observed in waves propagating along coronal tubes}

\begin{figure}[htbp]
\resizebox{9cm}{!}{\includegraphics{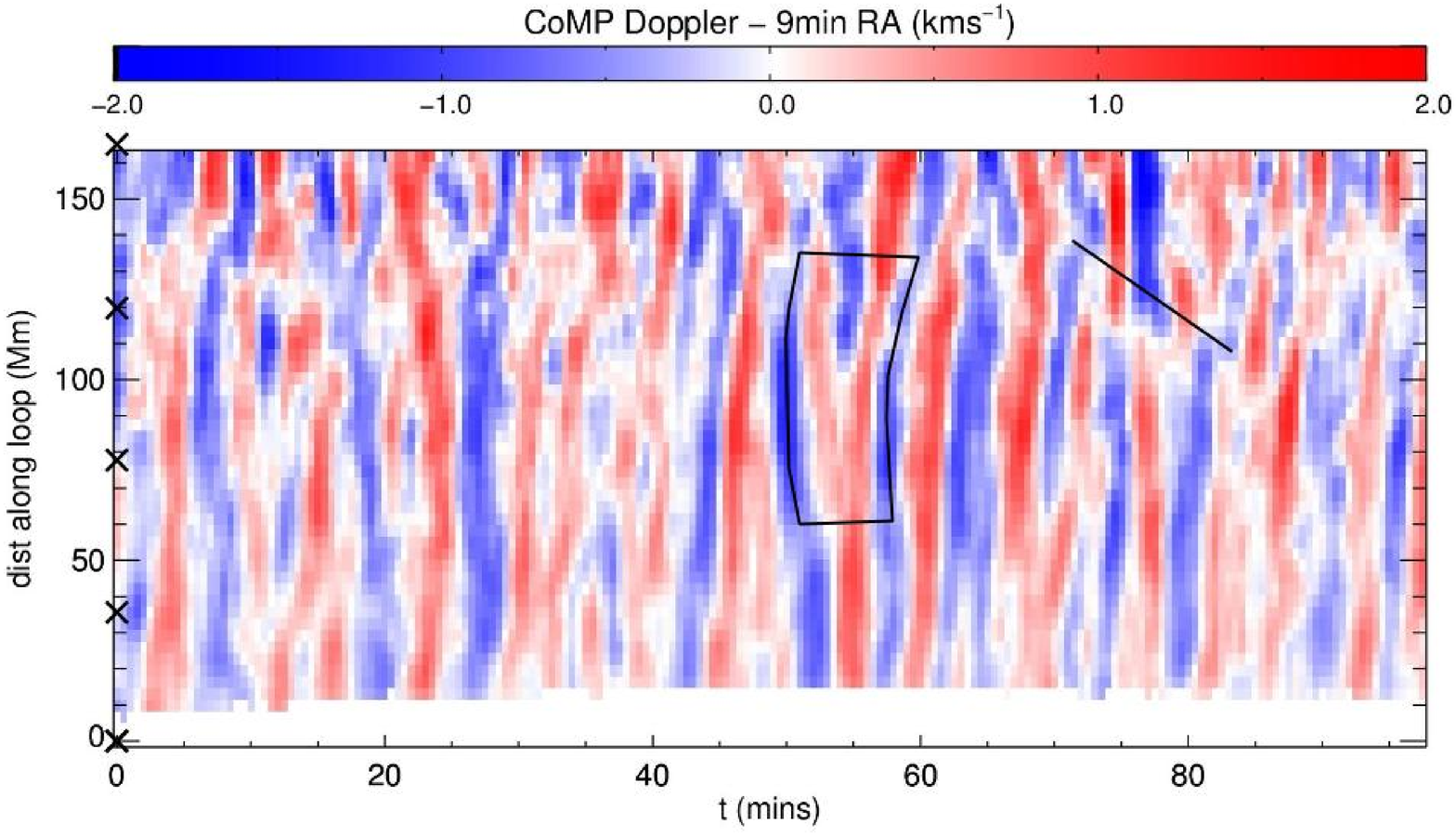}} 
\caption{Observation of coronal waves in the Doppler velocity of a coronal emission line by COMP, from \cite{threlfall_first_2013}. The diagram has time in abscissas
and the coordinate $z$, along the loop and parallel to the magnetic field, in ordinates. Among the many dislocations visible, two have been marked. The
closed curve
(that could represent the monodromy) encloses an edge dislocation, while the straight segment follows a possible gliding dislocation.}
\label{obs}
\end{figure}

 Figure \ref{obs}  shows a plot of measured Doppler velocities as a function of time for a particular trajectory in the corona, measured in a coronal emission
line by the Coronal Multi-channel Polarimeter \cite[CoMP,][]{tomczyk_instrument_2008}. The observations were presented by \cite{threlfall_first_2013} 
\citep[see also ][]{tomczyk_alfven_2007,tomczyk_time-distance_2009}. The parameters of the observation, off the solar disk, imply that we are
observing here the velocity
of the emitting plasma roughly transverse to the direction of the coronal magnetic field. The periodicity of the signal is evident. 
An average frequency $\omega$ can be estimated and we should use it throughout this paper to describe a \textit{carrier wave} whose amplitude and 
phase  may be locally modified. Another clear feature of the observed waves is 
the tilt of some of the crests and valleys, indication of a propagating wave along the coronal feature and along the magnetic field 
\citep{threlfall_first_2013}.  Several dislocations
are also visible, more easily after comparing them with the previous examples. We have marked one of them around minute 55, as in Fig.\ref{ejemplo}, 
with a closed curve made of 4 segments. A monodromy of interest would be the integration of the phase of the wave along this closed curve. 
Due to the importance, before 
proceeding, of classifying the observed wave pattern as a dislocation, we are going to compute the integral over the monodromy. Fig. \ref{obs} shows the actual Doppler 
shift or velocity of the plasma, not its phase. We cannot therefore integrate directly the measured values over the curve. In the Appendix we show 
how the  computation of the integral directly from the data can be made, but here we are going to take a more heuristic approach: One can always
deformate the curve so
that it is made of 4 segments along which we can safely identify the phase of the wave or its changes from the observations and integrate these
inferred values. The  first such segment follows the blue valley to the left
of the dislocation from roughly position 130 to position 50 in the ordinates. We will draw this segment as to follow the phase 0 (let us agree on 
that valleys (blue) are 
at phase 0 and crests (red) at phase $\pi$). Given the interpretation of these blue and red stripes as the valleys and crests of a wave, we should 
also agree in that roughly at the center of the referred valley there is a continuous line at phase 0 over which we draw the segment. A similar segment is drawn over the next
blue valley to the right, also along the 0 phase value. The integral $\int d\chi$ along either one of these two segments is 0 since there is no change in phase.
We next join the two segments with a straight horizontal line at ordinates position 50 from left to right. This straight horizontal segment goes from a point at phase 0 to the next 
red crest at phase $\pi$ and then ends in the next valley at phase $2\pi \equiv 0$. The integral $\int d\chi = 2\pi$ along this segment. Similarly, at ordinate point
130 we draw a horizontal and straight line from right to left joining the two vertical blue segments. This time the horizontal line starts at phase 0 and goes 
over two red crests and one blue valley before ending up in the final blue valley where we drew the segment. The integral is $\int d\chi = -2 \times 2\pi$, where
the minus sign comes from going from right to left, rather than in the other sense. The full integral along this closed curve is therefore
$$\oint_C d\chi = 2\pi - 4\pi=-2\pi,$$
different than zero. Hence, it has  a charge $m=-1$ given the chosen orientation of the curve.   A direct computation of the integral confirms the result, as seen in the Appendix. 
From the theory of functions of complex variable, a closed integral over a function of complex variables is non-zero when it encloses a number of poles or 
singularities larger than the amount of zeros. We conclude that somewhere inside this closed curve there is a singularity that result
in a nonzero, non trivial, monodromy.

This computation of the monodromy can be repeated for many other similar points of the figure, in particular for the line of dislocations marked with a
black 
tilted line that we identify as a possible gliding dislocation after comparison with Fig. \ref{ejemplo}. On the other hand, the main dislocation marked 
with the closed curve, and whose monodromy we computed, appears to be of the edge type. No pure vortex is seen. In order to explain these dislocations it is important
to stress that the axes of the plot are time $t$ and distance $z$ along a coronal loop as projected onto the plane of the sky. The edge dislocations seen in the figure 
are therefore localized at particular values of $t$ and $z$, but with freedom about the transverse coordinates $x,y$ or $r,\theta$ in the more appropriate cyclindric
reference system. 
Other assumed constraints beyond $z$ being along the loop  \citep{threlfall_first_2013} are that the magnetic field follows
the coronal loop and the wave propagates along this magnetic field. Since coronal loops appear as structures in coronal emission lines whose intensity variations
are  related to density enhancements, we further assume that the wave is propagating in a high-density cylinder.
The phase speed of the propagating wave, finally, has been measured to be in the range $700-1000$ km/s. Such speeds are much larger than the sound speed of the
corona and are comparable to the Alfv\'en and  kink speeds  characteristic of the propagation of Alfv\'en waves and fast magnetoacoustic body waves in coronal loops 
respectively.

\section{Interpretation of the observed dislocations in coronal waves}
A dislocation is a phase singularity. At the exact point of the phase singularity, the amplitude of the complex wave  has to be zero. This
statement allows us to search for dislocations as zeros of the complex amplitude of the wave. We want to describe a wave $\psi$ propagating at 
a characteristic frequency $\omega$ and wavenumber
$k$ along the $z$ axis, parallel to the magnetic field and the coronal loop. One common form for the solution of such a wave is written as
$$\psi=f(r,\theta,z,t)e^{i(kz-\omega t)}.$$
We notice that $f$ may depend on the three cylindric coordinates and on time, in the most general case. Since the term $e^{i(kz-\omega t)}$ can be neither
zero nor
singular it is obvious that all the information on the dislocations of this wave is contained in the $f$ function that should be complex in general.
We are therefore looking for those times and places $(r,\theta,z,t)$ where the complex amplitude 
$$\| f(r,\theta,z,t)\|=0.$$
A plane wave would have
$$ f(r,\theta,z,t) = A,$$ 
with $A$ real. This amplitude  cannot be zero, unless $A=0$. Plane waves carry no dislocations therefore.

All the analytical solutions found and given of magnetohydrodynamic waves in coronal loops
\citep{edwin_wave_1983,priest_solar_1982,roberts_wave_1981,goossens_surface_2012,goossens_nature_2009} can be written as
\begin{equation}
 \psi=f(r,\theta)e^{i(kz-\omega t)}
 \label{wave}
\end{equation}
with no dependence of $f$ on either $z$ or time. This simplification can be justified in the case of homogeneous loops along the magnetic field and
constant in time, at least
for periods long compared with the period of the wave. This assumption also makes possible to determine and fix $\omega$ and $k$. The actual form of the function
$f(r,\theta)$ changes with the assumptions and physical phenomena considered, but it always retains these dependencies. If such a wave carries a 
dislocation, where the phase is singular, it 
will be found at those points $(r,\theta)$ where the amplitude of the wave satisfies the condition
$$\|f(r,\theta)\|=0.$$
This equation will be valid for all values of $z$ and $t$. The dislocation therefore will be localized in the transverse plane $(r,\theta)$ but not in the 
plot $(z,t)$ of Fig. \ref{obs}. This is exactly the opposite situation to the one observed. The description of the wave as $f(r,\theta)e^{i(kz-\omega t)}$ cannot
produce a point dislocation in the plane $(z,t)$ and cannot therefore describe Fig. \ref{obs}, independently of the actual form of $f(r,\theta)$.
In view of this and since the observed dislocations are certainly localized in $(z,t)$, we may doubt about the description of the wave as given in
Eq. (\ref{wave}). 
Waves propagating along the field have to have this 
functional form as long as the assumptions made above on the loop hold, so we could as an alternative consider that the observed wave is propagating 
across the field. This 
possibility was however quickly discarded after inspection
of the possible analytical solutions to such a wave. We are not going to give here the details.

Another possibility we have inspected is that what we are observing is the interference of more than one wave. This possibility has been put forward by
\cite{threlfall_first_2013} and \cite{tomczyk_time-distance_2009} from actual 
inspection and Fourier filtering of the observations. Although it is unclear whether the methods used by those authors are still valid in the presence of phase singularities, 
let us follow them and suggest the possibility of two waves propagating with different frequencies $\omega_1$ and $\omega_2$. Let us further assume that 
the frequency difference $\Delta \omega= \omega_2 - \omega_1$ is small, \textbf{since \cite{threlfall_first_2013} filtered the observations in the Fourier space}. This wave can be written as
\begin{eqnarray}
 f(r,\theta)e^{i(kz-\omega_1 t)}+g(r,\theta)e^{i(kz-\omega_2 t+\alpha)}= \nonumber \\
 \left(f(r,\theta)+g(r,\theta)e^{-i\Delta \omega t+i\alpha}\right)e^{i(kz-\omega_1 t)}\approx \nonumber \\
 \left(f(r,\theta)+g(r,\theta)-g(r,\theta)i(\Delta \omega t-\alpha)\right)e^{i(kz-\omega_1 t)}
\end{eqnarray}
a complex amplitude multiplying the propagation term $e^{i(kz-\omega_1 t)}$.
Assuming for simplicity, but without loss of generality, that $f$ and $g$ are real amplitudes, we find that this wave can carry a dislocation when the two
conditions
\begin{eqnarray}
 f(r,\theta)+g(r,\theta)=0 \nonumber \\
 g(r,\theta) (\Delta \omega t-\alpha) =0
\end{eqnarray}
are simultaneously satisfied. Notice that the two equations are the real and imaginary part of the complex amplitude of the wave. If we require that both are
simultaneously zero, we get that the real amplitude of the wave is zero and that the phase is undefined, a dislocation.
The second one of those equations, the one for the imaginary part, is immediately satisfied at $t =-\alpha/\Delta \omega$ (modulo $\pi$) and so this wave may carry a dislocation 
localized in time at the position $(r,\theta)$ where the real part of the complex amplitude is also zero. It is 
sufficient that two waves, even plane waves, with slightly different frequencies interfere 
for a dislocation localized in time to be possible. In particular two waves with same amplitude and in antiphase $f=-g$ will always carry one dislocation at 
time
$t=\alpha/\Delta \omega$. Such dislocation would appear in observations like the one of Fig. \ref{obs} as a vertical line of dislocations. We do not
observe such dislocation. We further need to localize the dislocation at a particular position $z$ along the coronal loop.

A first attempt would be to suppose that the two interferring waves also have slightly different wavenumbers $k$. It can be easily seen that one of the conditions
for a dislocation in such a case would be
\begin{equation}
 i g(r,\theta) (-\Delta k z+\Delta \omega t-\alpha) =0
\end{equation}
The solution to this equation is a straight line $z=\frac{\Delta \omega}{\Delta k} t$ of dislocations with slope the ratio of differences in wavenumber and 
frequency. This is a encouraging result, since it could be the explanation of the tilted gliding dislocation marked with a black line in Fig. \ref{obs}. Although
in our explanation the line of dislocations is continuous, while in the observation is limited to roughly 3 periods, we could propose that we are seeing there
an interference of two waves with slightly different wavenumber and frequency, the ratio of which can be measured in the slope of the line, 
resulting in a dislocation that surfs the wave along the coronal loop. 

But such cannot be the explanation for the more common edge dislocations observed. We return to the general expression of Eq. (\ref{wave}) and we see that if 
$f(r,\theta)$ 
is complex, the condition of zero amplitude translates into both the real and imaginary parts being zero independently. If we require a dislocation to be
localized
in two coordinates, like z and t, both the imaginary and real parts have to depend on them. The interference of two waves with similar wavenumber and/or frequency
allows us to introduce an imaginary term $ i g(r,\theta) (-\Delta k z+\Delta \omega t)$ to the complex amplitude of the wave, but not to the real part. We end up 
therefore with either just one coordinate fixed for the dislocation or with a linear relationship between both, as we saw, but not with a complete determination
of both coordinates. To achieve this we will need the condition on the cancellation of the real part of the complex amplitude to involve $z$, $t$ or
both.  In the interference of the two waves proposed in this section, the other vanishing condition reads
$$ f(r,\theta)+g(r,\theta)=0, $$
which does not depend on either $z$ or $t$. The scenario of two interferring waves can at most explain one kind of observed dislocation, the gliding edge, but
not the most common one, the edge dislocation, in the observations of Fig. \ref{obs}.

\section{Simulating observed dislocations}

\begin{figure}[htbp]
\centerline{\resizebox{9cm}{!}{\includegraphics{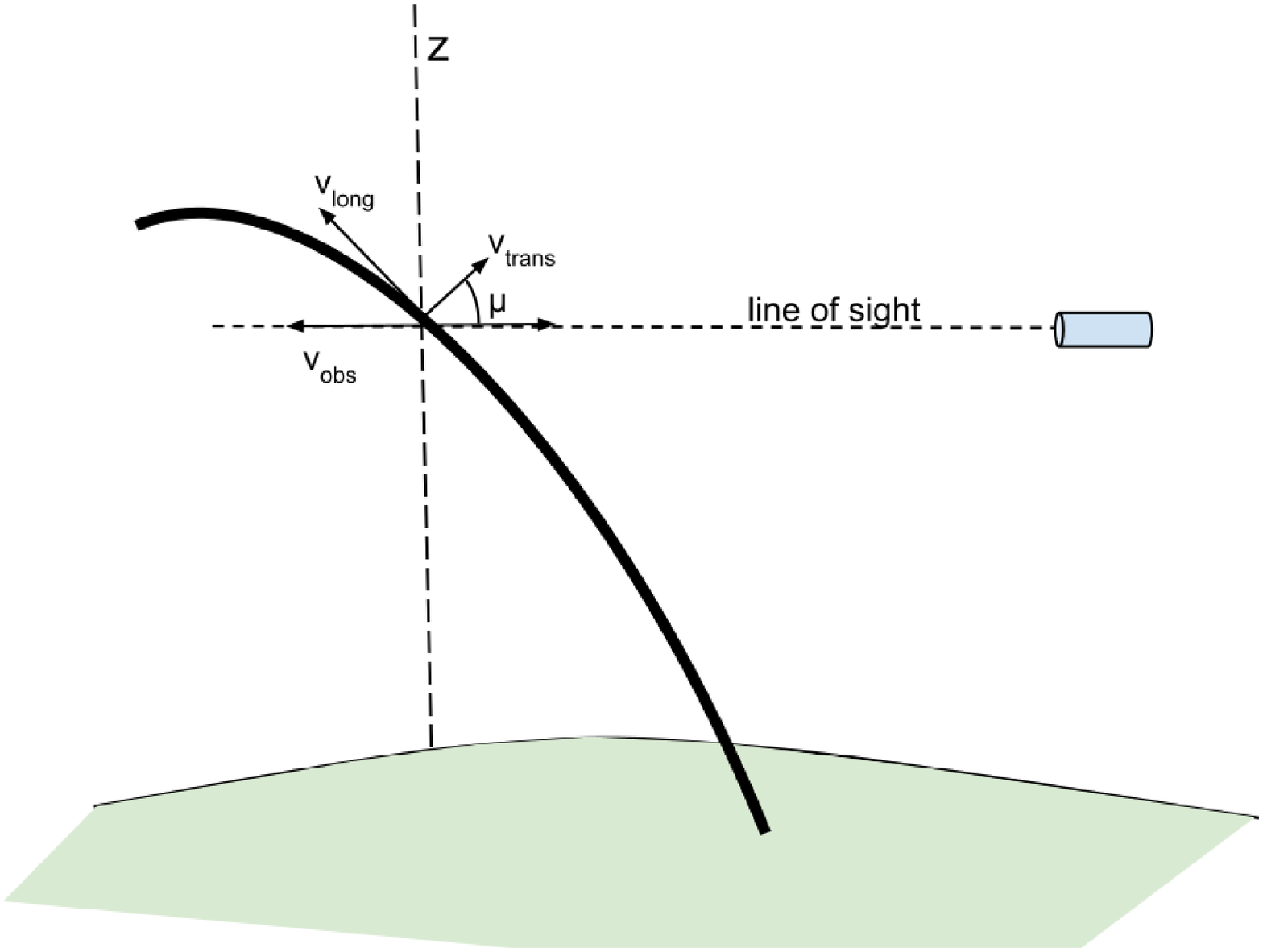}} }
\caption{Cartoon showing the loop, the transverse and longitudinal velocities of the waves and their compination into the observed velocity.}
\label{cartoon}
\end{figure}

Since magnetoacoustic waves with different velocities could explain gliding dislocations, we continue exploring this scenario of wave mixing. 
We now need to consider the fact, already mentioned, that the observed velocity is not necessarily the transverse velocity, but the combination of the projection 
of both the longitudinal
and the transverse velocities on the line of sight, since the loop is not on the plane of the sky (see cartoon in Fig. \ref{cartoon}).
In the case of 
Alfv\'en waves this makes no difference since the Alfv\'en wave has no longitudinal velocity component. But magnetoacoustic waves do have a longitudinal velocity
component. If the angle of projection is $\mu$, the appropriate velocity projected along the line of sight and observed in Fig. \ref{obs} is
$$ v_{los}=v_{long} \sin \mu(z) - v_{trans}\cos \mu(z)$$
where $v_{trans}$ is already the appropriate projection by the azimuthal angle, a projection to which we should come back later. In that expression,
we already
made a crucial change: the angle $\mu$ will change as we move along the coronal loop, since the loop is roughly a semi-circle starting and 
ending in the solar 
photosphere.  Therefore there is the potential of finding a dislocation at a given $z$ just because at that point the projection angle $\mu(z)$ is the appropriate
one. We could develop the conditions for such a dislocation to appear, but we rather notice first that this would give us a dislocation at a given $z$ for all
times, something which is not observed. Our next goal is to find a dislocation at fixed $z$ and $t$. Our solution is to mix the two mechanisms just proposed: 
The projection of the velocity over the line of sight that fixes $z$ and two waves propagating simultaneously but with different frequencies, what fixes $t$ of 
the dislocation. Those two simultaneously propagating waves are seen at a projection angle $\mu(z)$ so let us compute what would be seen. 

A first attempt can be made with two waves having the same value of $m$. This does not work. No dislocation can be observed in this configuration. The
second attempt 
concerns two waves of different value of $m$. This was already suggested by 
\cite{lopez_ariste_dislocations_2013} as a means to produce observed dislocations.  Let us suppose first that both waves are magnetoacoustic ones. We 
can suppose here
a mode with $m=1$, a kink, superposed with a mode with $m=0$, a sausage. This comes handy because the dispersion relations \citep{edwin_wave_1983} show that 
these two modes
propagate at different speeds and frequencies. The simultaneous existence of a sausage and a kink mode in the presence of a coronal density tube
will imply that, either the tube has a large enough radius, or that the sausage mode is a slow mode, two alternatives to which we should return later.
\textbf{To proceed with our description we need to have explicit expressions of the velocities of those wave modes. To write them we are going
to place ourselved in a simple model with a coronal loop formed by a uniform cylindric tube of plasma with piece-wise constant density. In the interior
of that cylinder, the  magnetoacoustic kink mode can be written as}
\begin{eqnarray}
 v_z&=&- i A_1 \frac{c_s^2k_1}{\omega_1^2} J_1 (m_1 r)e^{i\theta}e^{ik_1z-i\omega_1t}  \nonumber \\
  v_r&=&- A_1 \frac{\omega_1^2-k_1^2c_s^2}{\omega_1^2m_1^2}\left(J_0(m_1r)-\frac{1}{m_1r}J_1(m_1r) \right)e^{i\theta}e^{ik_1z-i\omega_1t} \nonumber  \\
  v_{\theta}&=&-iA_1\frac{\omega_1^2-k^2c_s^2}{\omega_1^2m_1^2}\frac{1}{r} J_1 (m_1 r)e^{i\theta}e^{ik_1z-i\omega_1t}   ,
  \label{Kink}
\end{eqnarray}
where $m_1$ is the $m_0$ defined by \cite{edwin_wave_1983} for the case of the kink mode frequency and wavenumber. \textbf{Always using the same loop model, 
and } with a similar redefinition of
$m_0$, the magnetoacoustic
sausage ($m=0$) 
mode can be written as
\begin{eqnarray}
 v_z&=&- i A_0 \frac{c_s^2k_0}{\omega_0^2} J_0 (m_0 r)e^{ik_0z-i\omega_0t}  \nonumber \\
  v_r&=&- A_0 \frac{\omega_0^2-k_0^2c_s^2}{\omega^2m_0^2}J_1(m_0r)e^{ik_0z-i\omega_0t} \nonumber  \\
  v_{\theta}&=&0  ,
  \label{Sausage}
\end{eqnarray}

The transverse velocity in both cases has yet to be projected  onto the plane that contains the line of sight and the $z$ direction. For this we have first to 
define the  direction $\theta=0$. Without loss of generality, we set this direction along the line of sight. This choice simplifies the expressions
above  and for the magnetoacoustic kink we can write
$$v_{trans}=A_1 \frac{\omega_1^2-k_1^2c_s^2}{\omega_1^2m_1^2}e^{ik_1z-i\omega_1t}\left[-m_1J_0(m_1r)(\cos^2\theta+i\sin\theta\cos\theta)+\right.$$
$$+\left. \frac{1}{r}J_1(m_1r)(\cos 2\theta+i\sin 2\theta)\right]$$
while in the case of the  magnetoacoustic sausage the transverse velocity is just
$$v_{trans}=A_0 \frac{\omega_0^2-k_0^2c_s^2}{\omega_0^2m_0^2}e^{ik_0z-i\omega_0t}J_1(m_0r)\cos\theta$$
Before combining the transverse and  longitudinal components projected onto the line of sight for both modes, and in the sake of reducing the
number of long  intermediate expressions, we are going to introduce the last ingredient of our model. The interference of two waves with 
different frequencies fixes the time of  occurrence
of the dislocation. The position $z$ of the dislocation is given by the position along the loop at which  the  longitudinal and 
the transverse components of both wave  modes are projected with the right angle. 
But, with the expressions we have at hand at this point, we realize that 
diferent points in
the plane $(r,\theta)$  place the dislocation at different values of $z$ and $t$.  To find a dislocation at $z$ and $t$ independently 
of $r$ and $\theta$, we are going to assume that the cross-section of the coronal loop is smaller than the spatial resolution of the CoMP instrument.
Coronal loops are mostly unresolved with present instruments, and certainly they
are so with  CoMP. We are therefore assuming that the full transverse plane $(r,\theta)$ of the wave is contained in one pixel, that is, that the 
radius $R$ of the coronal  tube is smaller than the pixel, and this forces us to integrate the velocities in both 
variables $r$ and $\theta$:
$$v_{obs}=\frac{1}{\pi R^2}\int_0^Rdr\int_0^{2\pi}rd\theta v_{los}.$$
The integral in $\theta$ is particularly interesting since it cancels out most
of the terms in the expressions of the velocity, symmetric in azimuth. This cancellation of unresolved velocities has also been pointed as a concern
for the  observation of pure Alfv\'en waves with $m=0$, which are torsional azimuthally symmetric waves. While the sausage mode has to be magnetoacoustic, we notice at 
this point that the kink mode could either be magnetoacoustic or Alfv\'en, both cases resulting in a non-zero transverse velocity after integration on $r$ and
$\theta$. In what follows we will pursue the calculations for the magnetoacoustic kink mode, keeping in mind that the magnetoacoustic sausage
mode could also be interferring with an Alfv\'en kink mode.

After integration on $r$ and $\theta$, the observed velocity from the magnetoacoustic kink mode, with both
the longitudinal and the transverse velocities combined, is
\begin{equation}
 v_{obs,1}=\frac{1}{ R^2} A_1 \frac{\omega^2-k_1^2c_s^2}{\omega_1^2m_1} e^{ik_1z-i\omega_1t}\cos \mu \int_0^RrJ_0(m_1r)dr
\end{equation}
while the sausage mode is seen as
\begin{equation}
 v_{obs,0}=-\frac{2}{ R^2} iA_0 \frac{k_0 c_s^2}{\omega_0^2} e^{ik_0z-i\omega_0t}\sin \mu \int_0^RrJ_0(m_0r)dr
\end{equation}

Finally, to combine the two waves we rewrite $\omega=\omega_1$  and $\omega_0=\omega_1+\Delta \omega$.  For simplicity, we will assume that $k_0=k_1=k$ since
this does not alter the results. The observed velocity, following
our model, will be
\begin{eqnarray}
 v_{obs}&=&e^{ikz-i\omega_0t}\frac{1}{R^2}\left[ \frac{\omega^2-k^2c_s^2}{\omega^2m_1}A_1\cos\mu\int_0^RrJ_0(m_1r)dr-\right.\nonumber \\
 &-&\left. 2 i\frac{k c_s^2}{(\omega+\Delta \omega)^2}A_0\sin\mu\cos \Delta\omega t\int_0^RrJ_0(m_0r)dr+\right.\nonumber\\
 &+& \left. 2\frac{k c_s^2}{(\omega+\Delta \omega)^2}A_0\sin\mu\sin\Delta\omega t\int_0^RrJ_0(m_0r)dr\right]
\end{eqnarray}
This observed velocity wave will show a dislocation when both the imaginary and real parts of the amplitude are simultaneously zero. This leads to the following
two equations
\begin{eqnarray}
  \frac{\omega^2-k^2c_s^2}{\omega^2m_1}A_1\cos\mu\int_0^RrJ_0(m_1r)dr+\nonumber \\
  +2\frac{k c_s^2}{(\omega+\Delta \omega)^2}A_0\sin\mu\sin\Delta\omega t\int_0^RrJ_0(m_0r)dr =0\\
  \frac{k c_s^2}{(\omega+\Delta \omega)^2} \sin\mu\cos \Delta\omega t =0
\end{eqnarray}
The second of these two conditions implies that, other than the trivial case $\mu=0$, the observed velocity will present a dislocation only if
$$\cos \Delta\omega t =0$$
that is, at those times when the two waves are in anti-phase\footnote{Had we used the Alfv\'en kink ($m=1$) mode instead, the condition here would have been that both
waves have to be in phase.}. This fixes, as expected, a time $t$ when the dislocation is possible depending on the difference
in frequency and on the precise moment when either wave was excited at, let us say, the feet of the loop. It is interesting to notice that,
given the wavelength of the coronal waves,
one does not expect this cophasing to happen more than once per loop. By inserting this cophasing condition in the first condition for the dislocation
we obtain a formula for the angle $\mu$ at which such dislocation would be visible:
\begin{equation}
\tan \mu=-\frac{1}{2}\frac{A_1}{A_0}\left(\frac{\omega^2}{c_s^2}-k^2\right)\frac{1}{m_1k}\left(1+\frac{\Delta \omega}{\omega}\right)^2
\frac{\int_0^RrJ_0(m_1r)dr}{\int_0^RrJ_0(m_0r)dr},
\label{tgmueq} 
\end{equation}
with an equivalent formula for the case of an Alfv\'en kink wave interferring with the magnetoacoustic sausage mode.

\begin{figure}[htbp]
\resizebox{9cm}{!}{\includegraphics{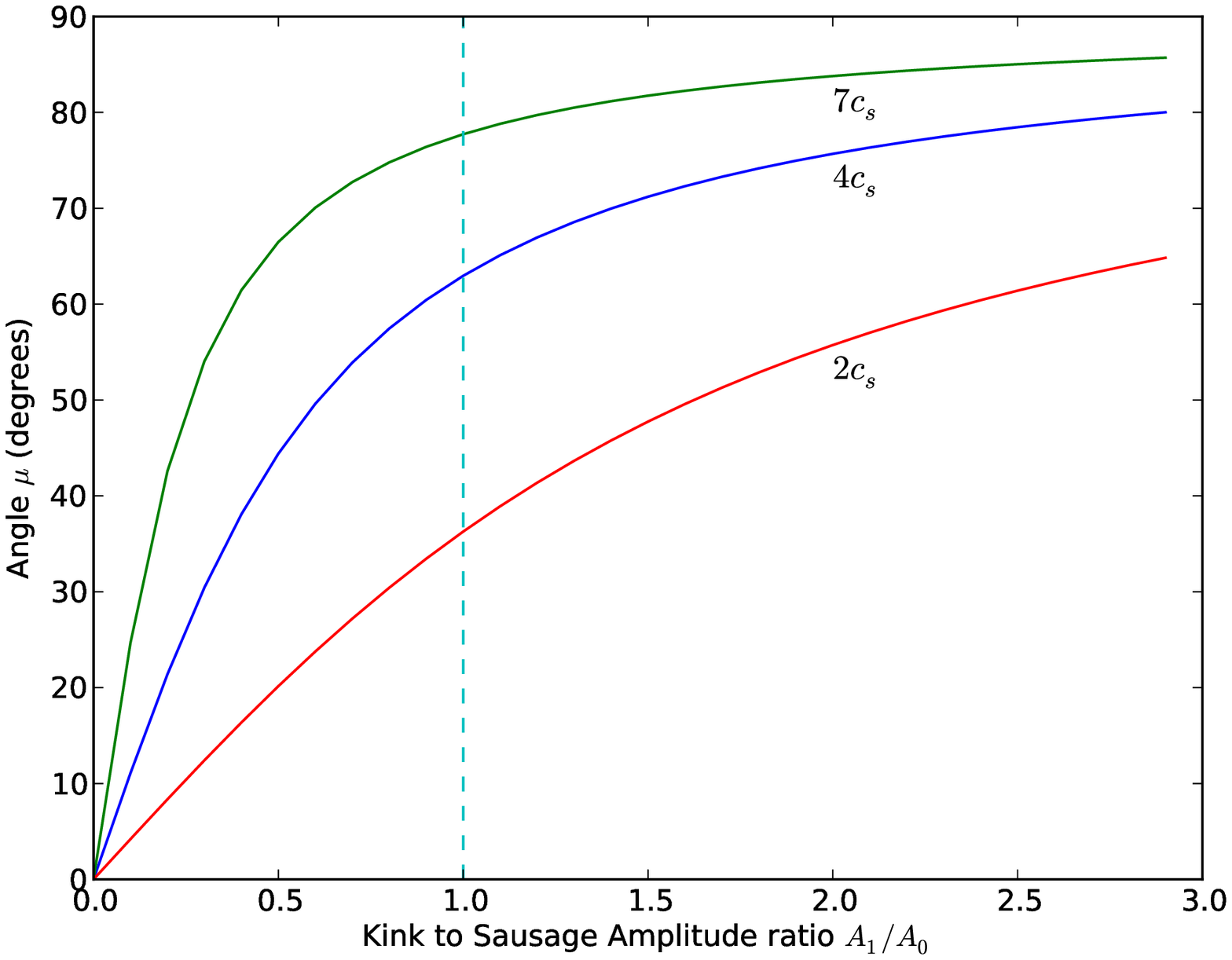}} 
\caption{Angle $\mu$ at which the dislocation will be seen following Eq. (\ref{tgmueq}) for different velocities of the magnetoacoustic kink wave, assuming a sound speed $c_s=100km/s$, 
an Alfv\'en speed  10 times the speed of sound, a period of 3 minutes for the kink wave and a 5\% difference in frequency between the kink and the sausage waves.}
\label{tgmu}
\end{figure}

\begin{figure}[htbp]
\resizebox{9cm}{!}{\includegraphics{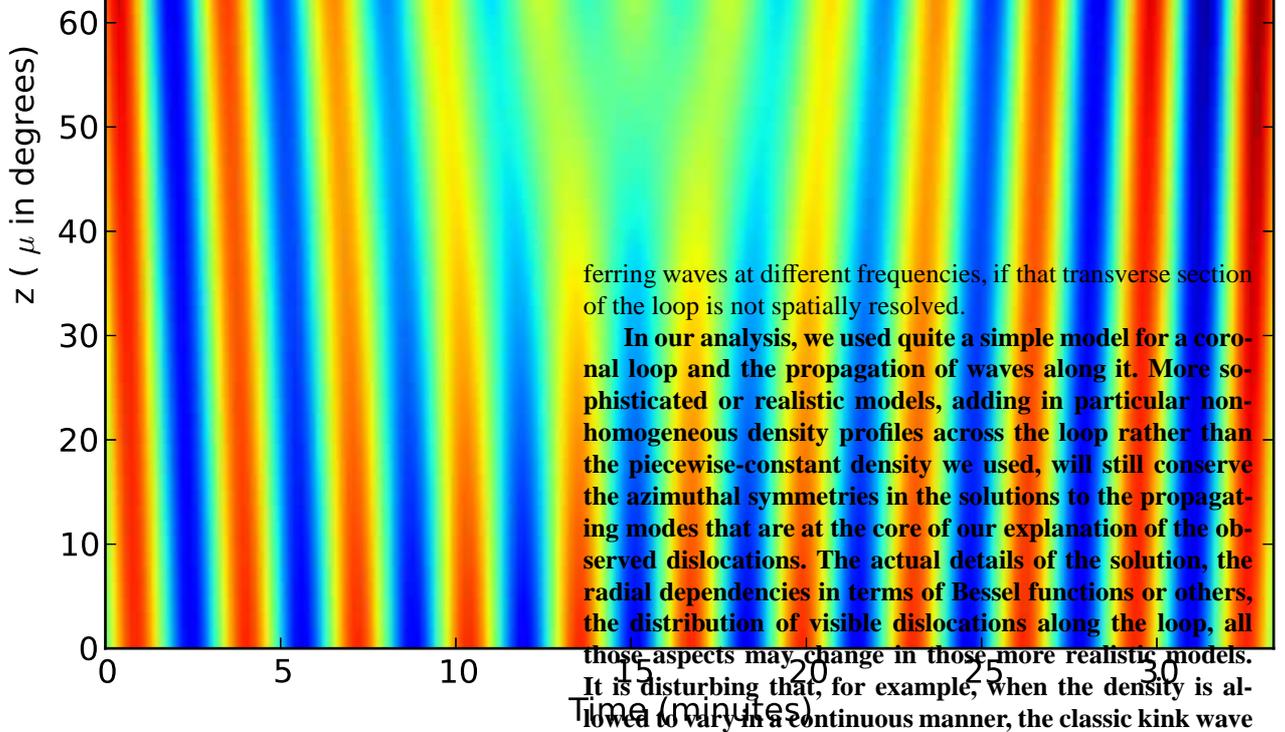}} 
\caption{Simulation of an observation of magnetoacoustic kink and sausage modes propagating along a circular loop with a 5\% difference in frequency, at a ratio of 7 in phase speeds
and a ratio of amplitudes $\frac{A_1}{A_0}=0.25$. The $z$ coordinate along the loop is given in terms of the angle $\mu$ it forms with the line of sight at each point.}
\label{simul}
\end{figure}

A dislocation will be seen at time $t$ at the position $z$ where the two waves happen to be in phase and which is 
 seen by the observer at an angle $\mu$ given by Eq. (\ref{tgmueq}). Fig. \ref{tgmu} shows 
examples of the variation of the angle $\mu$ with the ratio of amplitudes for 3 different values of the phase speed of the magnetoacoustic kink wave. It has been 
computed for
typical values of the sound and Alfv\'en speed in the corona (100 and 1000km/s respectively) for an average wave period of 3 minutes and a 5\% difference in 
frequency between the two waves. As an example of what would be seen, we show in Fig. \ref{simul} the observed Doppler velocity for the case of a 
fast kink wave propagating at 7 times the speed of sound (700 km/s), interferring with a slow sausage wave, that presents a ratio of 
amplitudes $\frac{A_1}{A_0}=0.25$. Notice that these
are the scalar amplitudes of the full vector wave or, in other words, the amplitudes of the pressure wave. The observed velocities have amplitudes given by these
$A_1$ and $A_0$ times other factors  (see Eqs. \ref{Kink} and \ref{Sausage}) resulting in the kink velocity having a larger amplitude than the sausage velocity even if $A_1$ is
smaller than $A_0$. With these numbers, as expected, a dislocation appears  at
roughly $\mu=50^{\circ}$. The result is visually striking as a correct reproduction of the edge dislocations observed in coronal waves by CoMP and seen in
Fig. \ref{obs}. 

If sausage and kink waves were excited in perfect phase matching at the loop feet, the dislocation would be visible periodically at most at one position along the 
loop (with a very long period $P=\frac{2\pi}{\Delta \omega}$).
The observed changes in position and time mean that the conditions of excitation of kinks and sausages in the loop feet vary permanently. Notice that,
if the kink wave 
propagated at roughly the speed of sound, the factor $\frac{\omega^2}{c_s^2}-k_1^2$ would be zero and the dislocation would only be found at low values of $z$. 
This is not the case: dislocations are seen for all values of $z$ in Fig. \ref{obs}. From this we have to conclude that the kink mode either is an Alfv\'en mode
or it is a fast wave moving at 
speeds comparable to the Alfv\'en speed. If the sausage mode is a slow mode, which can  
propagate independently of the radius of the loop, then it is easy to imagine that the dislocation arises from the excitation of a kink mode, Alfv\'en or magnetoacoustic,
that propagates at high speed along the loop and catches up with a sausage slow mode excited some time earlier. When this happens will depend on 
their respective velocities,
but also on the time between both excitations. Where this happens (if visible) will depend on the respective velocities, but also on the ratio of amplitudes of 
both
waves. \cite{threlfall_first_2013} measured  a phase speed around 700 km/s, roughly corresponding to 7 times the speed of sound. We see that at this 
high speed, the
spread of dislocations observed in Fig. \ref{obs} translates into magnetoacoustic kink amplitudes smaller than the sausage ones, although 
with a preference for 
dislocations to concentrate
in the top part of the loop. This appears to be the case in Fig. \ref{obs}.

\section{Discussion}

Observations of waves in the Doppler measurements of coronal emission lines by CoMP show wavefront dislocations. Several qualitative and quantitative arguments
have been made to ensure this identification of the wave singularity in the observed data. The conclusive identification of these singularities called 
dislocations
requires an explanation in terms of the waves expected to propagate in those coronal regions and their velocities projected along the line of sight,
seen as Doppler shifts in the coronal emission lines observed by CoMP.

We have provided such an explanation while explicitly discarding several other possibilities. We recall that the observations were projected  in a 
diagram $z-t$ where 
$z$ was the direction of propagation of the waves, also assumed to be the direction of the magnetic field. The dislocations were mostly 
of the edge type, meaning that the phase singularity was found at isolated points in the $z-t$ plane. The waves naturally propagating
along the magnetic field in the corona carry dislocations, but those dislocations are of the edge type only on the plane transverse to $z$, which in cylindric 
coordinates we can refer to as the plane $r-\theta$. These basic dislocations carried by the waves propagating along the magnetic field in coronal conditions
cannot explain the observations. Our first conclusion in this paper is therefore that the observations cannot be explained by single propagating waves of one type 
or the other: the presence of dislocations forces us to look for more elaborated scenarios of propagating waves.

Three ingredients are necessary to fix the dislocation in one single point of the $z-t$ diagram as observed. First, the interference of two waves with different
frequencies can fix the dislocation in time, but not in $z$. Second, the combination of transverse and longitudinal velocity components by projection onto the
line of sight, can fix the dislocation in $z$ but not in time, as long as the transverse and longitudinal components belong to waves of different mode $m$
(sausage and kink waves for example). Furthermore, the position $z$ of the dislocation varies for different points in the transverse plane $r-\theta$.
Third, the 
integration of the line emission over the transverse plane $r-\theta$ allows to fix a single $z$ and $t$ for the whole emitting 
region with two interferring waves at different frequencies, if that transverse section of the loop is not spatially resolved. 

\textbf{In our analysis, we used quite a simple model for a coronal loop and the propagation of waves along it. More sophisticated
or realistic models, adding in particular non-homogeneous density profiles across the loop rather than the piecewise-constant density we used,
will still conserve the azimuthal symmetries in the
solutions to the propagating modes that are at the core of our explanation of the observed dislocations. The actual details of the solution, the
radial dependencies in terms of Bessel functions or others, the distribution of visible dislocations along the loop, all those aspects may change
in those more realistic models.  It is disturbing that, for example, when the density is allowed to vary in a continuous manner, the classic kink wave 
becomes a surface Alfven wave with a strong Alfvenic character \cite{goossens_nature_2009,goossens_surface_2012} and the solutions display rapid 
variations in the radial and azimuthal components at the resonant layer with the consequence that the global motion is quickly damped.  It is difficult to foresee what the consequences 
would be for the existence and properties of dislocations as described in this study.
But we are confident that the need for interference of two modes with different azimuthal number $m$ at slightly different frequencies with 
longitudinal
and transverse velocities combined in the projection onto the line of sight will still be the ingredients to explain the observed dislocations. Simpler
conditions do not seem to work, and more complicated ones may not be so common and harder to be simultaneously met.}

\textbf{Indeed, we appreciate in our explanation that} the three ingredients (interference of waves of different frequency, combination of longitudinal and transverse velocity components into the line of sight, and 
integration of the full emitting region inside the pixel) are quite natural and common. It is not surprising therefore that the observations show a 
large number
of dislocations during the observation. But the same three ingredients also limit the type of waves responsible for the observations. Thus the integration
over the $r-\theta$ plane excludes from the model any wave with azimuthal symmetry. Pure Alfv\'en waves with $m=0$, in particular, result in a zero 
signal and  can be  excluded. The need for a longitudinal velocity component implies the presence of at least one magnetoacoustic wave. Sausage modes
$(m=0)$  can be responsible of the longitudinal velocity component, but their transverse velocity is azimuthally symmetric and cancels out.
 Kink waves ($m=1$), either magnetoacoustic or Alfv\'en, can be responsible of the transverse velocity component, but not of the longitudinal one, in one case
 because it cancels out, and in the other because there is no longitudinal component.
The addition of a magnetoacoustic sausage plus a  kink wave, either magnetoacoustic or Alfv\'en, with slightly different frequencies, has all the
required 
properties. Integrated
across the loop, the observed Doppler shift is made of the transverse velocity of the kink mode plus the longitudinal velocity of the sausage mode. At
those points along the loop with the right projection angle, the interference of the two waves with the right phases results in a dislocation 
localized in $z-t$ as observed. The equation for the appearance of the dislocation also suggests that, at high speeds of the kink wave, dislocations will be more frequently observed
on the top  of the loops, but some variability in position is expected if the amplitude of the kink mode is  smaller than that of the sausage mode.
All this appears to be the case in the solar corona, and therefore we conclude that a fast kink mode, either a fast magnetoacoustic or and Alfvén one, catches up a slow sausage 
mode at some point along the loop and produces 
a dislocation visible with COMP. The fast character of the kink mode is suggested by the measured speed of the observed wave \citep{threlfall_first_2013}, but
also by the
comparison of the predictions of Fig. \ref{tgmu} with the observed positions of the dislocations in Fig. \ref{obs}. The slow character of the sausage mode makes it 
compatible with a propagation with cross-sections smaller than CoMP spatial resolutions that justify our integration in $r$ and $\theta$, and apparently smaller 
than the cutoff of fast sausage modes. A slow
sausage mode makes it also a good candidate to be over-run often by fast and/or Alfv\'en kink modes excited at different times.

\cite{threlfall_first_2013} compared the observed waves in CoMP with simultaneous observations of emission, and hence density, perturbations observed by AIA. 
The density
perturbation corresponding to our two waves is 
\begin{equation}
 \rho_1=-iA_0 \frac{\rho_0}{\omega} J_m (m_0 r)e^{ikz-i\omega t}
\end{equation}
Despite the absence of any azimuthal dependence, the radial integral of the $J_1(m_0 r)$ in the case of a fast magnetoacoustic kink is going to be almost 
negligible compared to the same integral 
of the $J_0(m_0r)$ function for the slow sausage mode. The compressional wave will therefore be dominated by the sausage mode. The Doppler signal,
on the other hand, will be dominated by the kink mode. Hence, the two instruments are sensitive to one or the other wave but not to both, and one
should not expect any
spatial correlation between observations of the compression wave by AIA/SDO and observations of the velocity wave by COMP. 
In spite of this, the observed periods (frequencies)
will of course be similar. This coincides with the conclusions of \cite{threlfall_first_2013}.

\section{Conclusion}

We have identified wavefront dislocations in the the observations of coronal waves made by CoMP. Explaining the observed dislocations has forced us to abandon
the image of a single wave propagating in those coronal structures. We explain in detail the only
scenario we have found that can explain the observations. Our model is made of two propagating waves with different wave frequencies and in interference. The two
waves have different azimuth dependence (or charge) and the observations integrate the velocity wave all over the cross-section of the wave, smaller than 
the spatial resolution of the instrument. This  eliminates many possible candidates for which  the signals cancel out after  integration. In particular torsional
Alfv\'en waves with $m=0$ are excluded but magnetoacoustic sausage waves appear as a necessity, combined with a kink mode which can be either a fast magnetoacoustic
mode or an Alfv\'en wave. The two wave modes, the magnetoacoustic sausage and the kink, propagating at different 
frequencies and integrated over the loop cross-section are seen under different projection angles at different positions along the loop. This projection
of the transverse and longitudinal velocities of the two waves onto the line-of-sight fixes when and where the dislocation will be seen.

Computation of these conditions for the visibility of the dislocations leads us to conclude that our model can reproduce the observations if we assume a fast 
kink mode (magnetoacoustic or Alfv\'en) catching up over a slow sausage mode propagating along the coronal loop. The observed signature is
dominated by the 
velocity amplitude of the 
fast kink mode although the pressure amplitude of the slow sausage is still larger than that of the kink mode. The observed dislocations also
imply, following
our model, that the spatial resolution of the observations is not enough to resolve the cross-section of the loop: we see an integrated signal.
\textbf{Despite the simplified model used for our analysis, the conditions under which dislocations can be seen in the data appear quite general and
translatable to more sophisticated models.}

These sausage modes required in our scenario to explain the dislocation may coincide  with the density waves observed by AIA/SDO \citep{threlfall_first_2013}. The 
density perturbation in our 
model is dominated by the sausage mode, even if the kink mode is magnetoacoustic. Observations of emission changes, mostly proportional to the density, would 
then mostly see this mode, rather than the 
kink one. Although of similar frequency and period, the very different wavelengths will make them difficult to correlate with the observed velocity wave, mostly
dominated by the kink mode.

Finally, we insist in that the observations cannot be explained with Alfv\'en waves alone, but require a combination of at least one magnetoacoustic mode. The 
relative  amplitude of these  modes will have to be explained by the excitation of these waves or its propagation and eventual dissipation in the low parts of 
the corona. The existence of other propagating modes cannot
be fully discarded, but since they cannot be responsible for the observations due, as for the case of the torsional Alfv\'en wave, to their azimuthal symmetries,
they have to be considered as non-observed or without relevance for the present observations. On the other hand, the continuous existence and interaction of 
two waves must be seen as a permanent feature of coronal loops.

\appendix
\section{Direct computation of the monodromy}
In the main body of this paper we computed the monodromy through the smart trick of identifying in the data places of known phase and drawing the closed
curve through those places. We drew lines along contiguous places of known phase in the crest or the valleys of the waves at both sides of the suspected
dislocation and joined them with straight line that crossed an integer amount of crest or valleys. This method was suggested by the organized patterns in the
data that allowed an easy and safe identification of those places of known phase. But in order to feel fully confident about the method and the presence of the
dislocation one would prefer to be able to directly solve the monodromy integral along any closed curve in a general wave field.

The observed data in our present case is the Doppler shift of an emission line interpreted as the line-of-sight velocity of the plasma. This is a real quantity.
Our first step will be to interpret these observations as the real part of a complex wave field of which we have to determine the imaginary part. 
Let us for simplicity assume  that  the observation can be safely interpreted as due to a wave with a unique average frequency $\omega$.
We can describe the observed wave field as follows
$$ \phi(z,t)= A(z,t)\cos\left(\omega t+\alpha(z,t)\right)$$
The observations, the real quantity $\phi$ at each position $(z,t)$, are described as a variable real amplitude $A(z,t)$ times a cosine variation in 
time with frequency $\omega$. We asume a constant zero time for the full wave field, but allow for a local phase shift $\alpha(z,t)$ which, through its time 
dependence, may include local frequency changes. The combination of the 
variable amplitude and local phase shifts allows the description of very complicated wave patterns, including the one in Fig. \ref{obs}, as long as one accepts
the constant average frequency $\omega$ over the time and place of the observation.

We can decompose the cosine function as
$$\cos \left(\omega t+\alpha(z,t)\right) =\cos \omega t\cos \alpha - \sin \omega t \sin \alpha$$
The local phase shift $\alpha$ can now be interpreted as a local modification of  the amplitude of two different waves:
$$\phi=\left[A(z,t)\cos \alpha \cos \omega t\right]-\left[A(z,t)\sin\alpha \sin \omega t\right]= $$
$$=\psi_C \cos \omega t +\psi_S \sin \omega t$$
This suggests the construction of the complex wave field
$$\psi(z,t)=(\psi_C+i\psi_S)e^{i\omega t}$$
This wave has an amplitude
$$ \psi_C^2+\psi_S^2 = A^2$$
the amplitude of the observed wave, while locally it adds a phase $\chi$ 
$$\tan \chi = \frac{\psi_S}{\psi_C}=\frac{\sin \alpha}{\cos \alpha}=\tan \alpha$$
identical to the local phase of the observed wave. Thus the proposed complex wave field
$$\psi(z,t)=A(z,t)e^{i\alpha(z,t)}e^{i\omega t}$$
has  the same observable parameters as the original real wave $\phi(z,t)$ and can be used instead of it,
with a straightforward (diffeomorphic) correspondence between them.

Since $z$ and $t$ are the coordinates of Fig. \ref{obs}, let us re-write this complex field for the wave on the longitudinal velocity as just a real amplitude
and a phase
$$ \psi(z,t) = \rho e^{i \chi}. $$
After differentiating this expression, we find that
\begin{equation}
d \psi = d \rho ~e^{i \chi} + i \rho~e^{i \chi} ~d \chi.
\end{equation}
Dividing by $\psi$
\begin{equation}
\frac{d \psi}{\psi} = \frac{d \rho}{\rho} + i ~d \chi ~.
\end{equation}
We  can integrate  both sides of this last expression along a closed curve $C$, the monodromy:
$$\oint_C \frac{d \psi}{\psi} = \oint_C \frac{d \rho}{\rho} + i \oint_C ~d \chi ~.$$
Since $\rho$, the amplitude of the complex wave, is, by definition, a real quantity we have that
$\oint_C \frac{d \rho}{\rho} = 0$
on any closed curve, and the monodromy simplifies to 
\begin{equation}
\oint_C \frac{d \psi}{\psi} = i ~\oint_C d \chi ~.
\end{equation}
The right part is the monodromy over the phase of the wave which, if different than zero, identifies the presence of a singularity, a dislocation, inside the 
closed curve.  The left part is an integral over the observed data. We conclude that from the observations we can built de complex field 
$\psi$ and then compute the integral on the left along the chosen closed path 
$C$ to obtain the required integral over the phase at right. This solves the problem of computing the monodromy on the phase directly from the data. 
It is 
illustrating to make one further step. The integral on the left side can be formally integrated
\begin{equation}\label{eq:connection-phase-value}
\log \frac{\psi_f}{\psi_i} = i ~\oint_C d \chi ~,
\end{equation}
where $\psi_f$ and $\psi_i$ are the final and initial values respectively of $\psi$ at the closed path. These would be the same, since the path is closed, but
for the $\psi$ function being complex. This opens the possibility of those initial and final values not being in the same Riemann surface of the logarithm 
function. Indeed the complex logarithm\footnote{The complex exponential function is not injective.} is $\log z = (\log z)_\mathrm{principal} + i 2 \pi N$. The 
principal part of the logarithm is identical for the initial and final points, so that we can conclude that
\begin{equation}
\oint_C d \chi = - 2\pi N,
\end{equation}
where $N$ is the number of twists made by the logarithm function as it follows the path $C$.

\begin{acknowledgements}
The authors  acknowledge financial support from the Spanish Ministry of Economy and Competitiveness (MINECO) AYA2011-24808, AYA2011-22846, and AYA2010-18029, 
from the Ram\'on y Cajal fellowship and from the projects CSD2007-00050 and ERC-2011-StG 277829-SPIA Starting Grant, financed by the European Research Council.
\end{acknowledgements}

\bibliographystyle{/home/arturo/TeX/aanda/bibtex/aa}
\bibliography{/home/arturo/0art/art50/art50.bib}

\end{document}